\documentclass [pra,noeprint,reprint,groupedaddress,showpacs]{revtex4-1}
\usepackage{graphicx}
\usepackage{dcolumn}
\usepackage{amsmath}
\usepackage{amsfonts}
\usepackage{amssymb}
\usepackage{bm}
\usepackage{nicefrac}
\usepackage{color}
\usepackage{float}
\usepackage{sidecap}
\graphicspath{{figures/}}
\usepackage{hyperref}
\usepackage[none]{hyphenat}
\hypersetup{backref,
pdfstartview=Fit,
colorlinks=true,
citecolor=blue,
urlcolor = blue,
linkcolor = blue}

\begin{document}
\title{Observation of the Magneto-Optic Voigt Effect in a Paramagnetic Diamond Membrane}

\author{Haitham A.R. El-Ella}
\email{haitham.el@gmail.com}
\author{Kristian H. Rasmussen}
\author{Alexander Huck}
\author{Ulrik L. Andersen}
\author{Ilya P. Radko}

\affiliation{Centre for Macroscopic Quantum States (bigQ), Department of Physics, Technical University of Denmark, 2800 Kongens Lyngby, Denmark}

	\begin{abstract}
	The magneto-optic Voigt effect is observed in a synthetic diamond membrane with a substitutional nitrogen defect concentration in the order of 200 ppm and a nitrogen-vacancy defect sub-ensemble generated through neutron irradiation and annealing. The measured polarisation rotation in the reflected light is observed to be quadratically proportional to the applied magnetic field and to the incident reflection angle. Additionally, it is observed to be modifiable by illuminating the diamond with a 532 nm laser. Spectral analysis of the reflected light under 532 nm illumination shows a slow narrowing of the spectral distribution, indicating a small increase in the overall magnetisation, as opposed to magnetisation degradation caused by heating. Further analysis of the optical power dependence suggest this may be related to a shift in the spin ensembles charge state equilibrium and, by extension, the resulting ensemble magnetisation. 
	\end{abstract}
\maketitle
\section{Introduction}
	Although diamond crystals should be considered as optically isotropic due to their cubic structure, the inherent presence of strain and defects results in an unavoidable degree of birefringence \cite{LANG1967}. While this may be of trivial origin such as from inherent growth discontinuities and the resulting crystal strain (and usually inconvenient e.g. \cite{Loon2006}), it is compelling when originating from magnetically active defects, with the potential to be of practical use. Such defects give rise to magnetic-field dependent birefringence and dichroism, which display a distinct inter-dependence between the defect-perturbed crystal symmetry, the defects intrinsic electronic ensemble properties, and an applied magnetic field \cite{Zvezdin1997}. Here we report on the experimental observation of magnetically dependent birefringence, also known as the magneto-optic Voigt effect (MOVE) \cite{Freiser1968}, in a synthetic diamond membrane with a high substitutional nitrogen (P1) concentration, and a fractional nitrogen-vacancy (N-$V^-$) sub-ensemble concentration.  
	
	Diamond-based spin ensembles, in particular N-$V^-$ ensembles, are compelling systems for the development of biologically compatible magnetometers \cite{Barry2016a}, masers \cite{Breeze2018}, and the exploration of novel collective quantum phases \cite{Choi2017}. Studying the magneto-optical aspects and origins of such ensembles is therefore pertinent for their continued development, and the observation of MOVE in such paramagnetic diamond systems is noteworthy due to the distinct optical spin-polarisation mechanism of the N-$V^-$ spin system \cite{Robledo2011, Doherty2013}, and in turn, other optically un-addressable magnetic spins that couple to the N-$V^-$ spins \cite{Purser2019, Alfasi2019}. While the observation of optical birefringence and dichroism in natural diamonds with high defect concentrations is well documented e.g. \cite{Konstantinova2006}, there is yet no reported observation, to our knowledge, of magnetically dependent and laser-induced birefringence in artificial diamonds. This presents the possibility of employing these systems for polarimetry-based magnetometry, while also suggesting a possible degree of intrinsic magnetic ordering that is of interest fir exploring collective quantum magnetic behaviour \cite{El-Ella2019} 
	
	The MOV effect, which is often encountered in the literature under the general term of linear magnetic birefringence, can be phenomenologically described as the presence of two orthogonal complex indices of refraction for an in-plane magnetised sample \cite{McCord2015}. The difference between these two orthogonal indices originate from the presence of spin-orbit and Zeeman spin-exchange energies which accompany an anisotropic magnetic dipole, such as that of the N-$V^-$ defect \cite{Doherty2012}, and its anisotropic coupling with neighbouring P1 electron spins \cite{Hall2016}. In either transmission or reflection geometries, this results in a change of the incident light polarisation through either polarisation-selective refraction or absorption (or both), and depends non-linearly on the light frequency, and quadratically on the applied magnetic field strength. The degree of polarisation change therefore depends on the relative orientation of the external magnetic field with the polarisation angle and crystallographic symmetry axis that defines the degree of anisotropy. 
	
		\begin{SCfigure*}
		\centering
		\includegraphics[scale = 0.65]{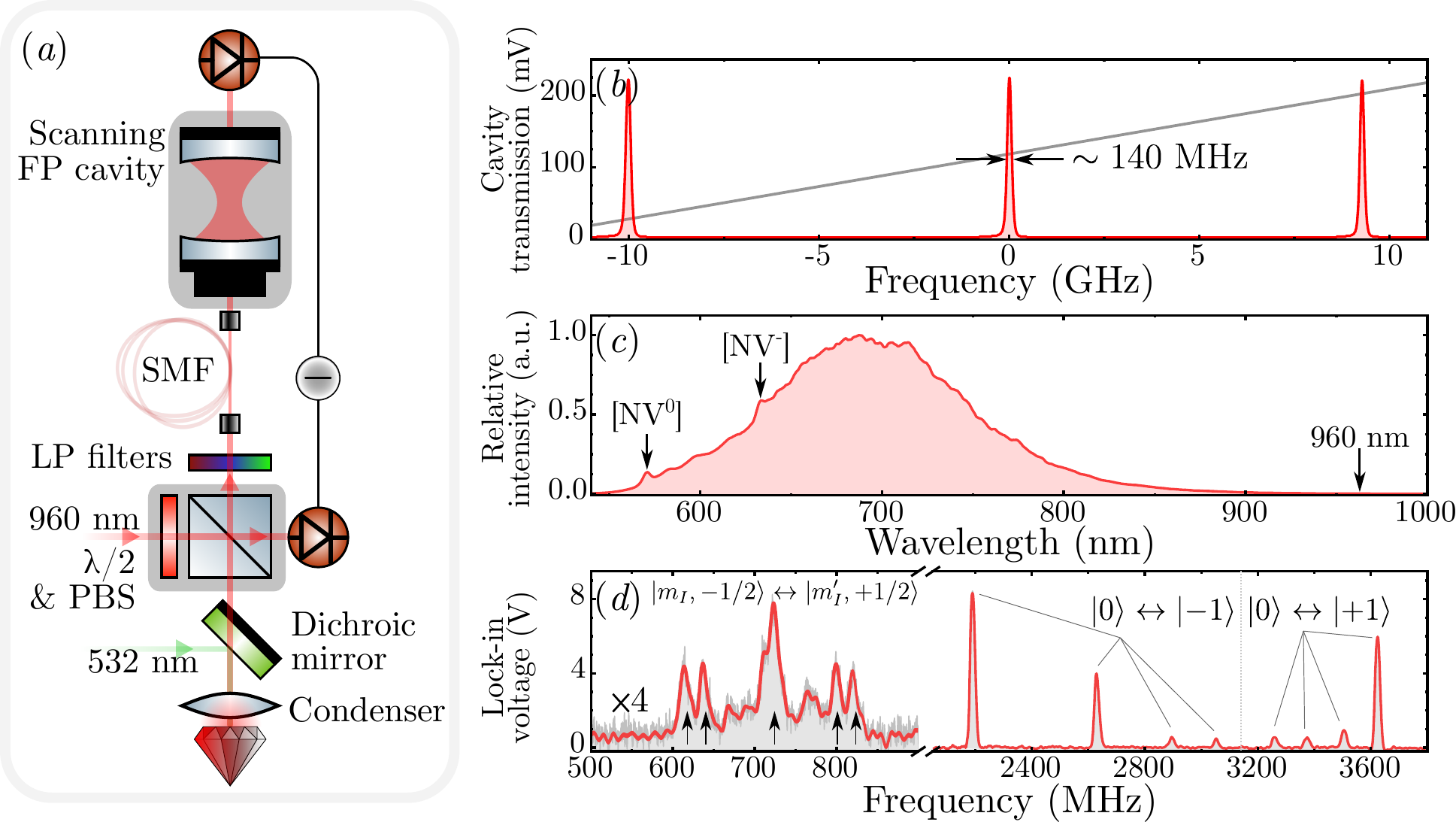}
		\caption{(\textit{a}) Summarised schematic of the experimental setup with abbreviations noted for the Fabry-P\'{e}rot (FP) cavity, a single mode fibre (SMF), long-pass filters (LP), a half-wave plate ($\lambda/2$) and a polarising beam splitter (PBS). Basic characterisation of the cavity is shown in (\textit{b}) with the grey line plotting the associated linear piezo-driving voltage ramp. A confocal optical and magnetic resonance spectrum are shown in (\textit{c}) and (\textit{d}), for which the optical contrast for magnetic resonance measurements was in the order of 1\% at -5 dBm applied microwave power.}
	\end{SCfigure*}
	
	Here we experimentally show a quadratic magnetic dependence of polarisation rotation in the light reflected from the surface of a paramagnetic diamond with a high concentration of magnetically active defects, as well as a quadratic dependence on its reflection angle while an in-plane magnetic field is applied orthogonal to the reflection plane, and obliquely to the N-$V^-$ symmetry axis. Intriguingly, the polarisation rotation is shown to be modifiable by illuminating the diamond using a 532 nm laser. Finally, the collected reflected and scattered light is spectrally analysed with the aim of outlining the mechanism underlying this observation.  

	\section{Experimental Setup and Diamond Characteristics}

	The employed experimental setup is summarily outlined in Fig.1(a), and is based on a confocal geometry which collects angular-resolved and polarisation-dependent reflected light, as well as optically detecting magnetic resonance (ODMR). The probe light to be reflected is passed through a $\lambda/2$ plate and a polarising beam splitter (PBS) cube, before passing through a dichroic mirror with a cut-off wavelength at 540 nm. The light is then reflected by 90 degrees onto an attached aspheric condenser lens (NA $\approx0.8$) used here as the objective lens. This aspheric objective focuses the linearly polarised light onto the surface of the diamond into a $\sim$2 $\mu$m diameter spot, and collects both reflected and scattered light, sending them back through the same path to the initial PBS. Only the 90 degree reflector and the objective lens rotate relative to the diamond sample and the $\lambda/2$ plate and PBS, with all the latter being fixed.
	
	This collected light exiting the PBS is first passed through two longpass filters with a 550 nm cut-off, and then either detected directly using a biased Si photo-diode, or focused into a single-mode fibre and then out-coupled, collimated, and focused into a scanning confocal Fabry-P\'{e}rot (FP) cavity with a free spectral range of approximately 10 GHz and a full width at half maximum-limited resolution of approximately 140 MHz (finesse $\approx70$), whose spectrum is shown in Fig.1(b). A probe field wavelength of 960 nm is used due to the negligible absorption at room temperature of both the triplet and singlet transitions for both N-$V^{-/0}$ charge states \cite{Kehayias2013}, as well as experimental convenience. 
	
	Subtraction of power-related noise inherent to the laser is performed by detecting and subtracting the light transmitted in the adjacent face of the PBS using an identically biased Si photo-diode. This light is optically balanced with the reflection from the diamond surface using neutral density filters, for a chosen relative configuration of the $\lambda/2$ plate and PBS. The experimental configuration thereby measures the change in polarisation of the reflected light through monitoring the difference in amplitude of the two beams transmitted through the PBS, ensuring that any measured trend is due to a relative change in polarisation rather than fluctuations of the laser power. However, this configuration does not remove the inherent fluctuation of the laser's polarisation, which was accounted for in the subsequent analysis. Furthermore, the reflected amplitude of the probe beam prior to the PBS was confirmed to be insensitive to variations in the applied magnetic field. 
	     
	\begin{SCfigure*}
		\includegraphics[scale = 1.1]{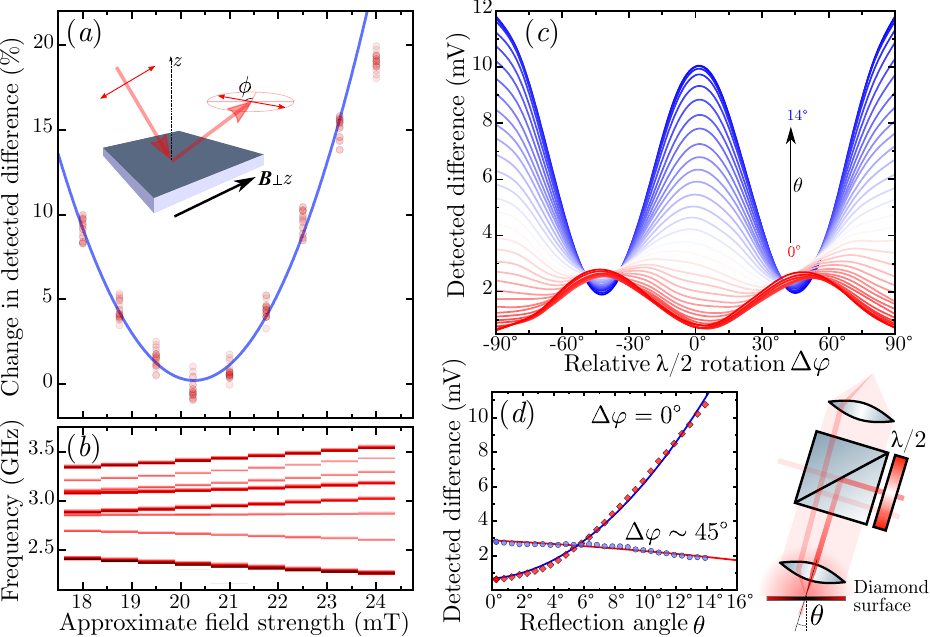}
		\centering
		\caption{(\textit{a}) Dependence of the reflected light polarisation as a function of applied magnetic field strength. The top graph shows the change in the difference between the detected amplitudes after the PBS, signifying a partial $\phi$ rotation of the incident linearly polarised light, and an overall elliptic polarisation. Each point in the top graph represent an average of 2500 successive measurements taken at a 4 kHz rate. A full data set at each magnetic field corresponds to 12 seconds of accumulation. (\textit{b}) A map of the ODMR spectra [\textit{cf}. Fig.1(\textit{d})] as a function of magnet field strength as the magnet is brought closer to the diamond. (\textit{c}) Dependence of the detected difference as a function of the relative rotation of the $\lambda/2$ plate by $\Delta\varphi$, for gradually increasing reflection angles $\theta$ from $0^\circ$ to $14^\circ$. (\textit{d}) A gradual inversion of detected intensity is observed alongside a quadratic increase in detected amplitude as $\theta$ is increased.}
	\end{SCfigure*}

	The diamond, similar to that reported in \cite{El-Ella2019a}, consists of a mechanically thinned and polished 1b diamond crystal (Element Six), with a \{100\} face, $\langle 100 \rangle$ edges, approximate dimensions of $50~\mu\text{m}\times9~\text{mm}^2$, and a specified P1 concentration in the order of 200 ppm. This was subjected to thermal neutron irradiation with an approximate fluence of $10^{19}$ cm$^{-2}$, followed by annealing at 900 $^\circ$C. A minimum, lower-bound N-$V^-$ concentration was estimated using confocal microscopy to be in the order of 0.5 ppm, however due to the dominant P1 concentration, significant quenching of the N-$V^-$ fluorescence is expected due to P1 optical absorption and N-$V^-$-P1 dipole coupling \cite{GattoMonticone2013}. The N-$V^-$ concentration is therefore expected to be at least in the order of 10 ppm. A metallic coplanar wave-guide designed for delivering microwaves with frequencies up to 20 GHz was deposited directly onto one of the polished diamond faces, consisting of 5 nm Ti and 200 nm of Au. The complete structure was wire-bonded to a co-planar wave-guide microwave board, with an optical aperture on the opposite side to facilitate illumination and reflection from the diamonds un-coated side. 
	
	The diamond is placed in the focal point of the objective, and a cylindrical rare earth magnet fixed to a translation stage is placed adjacently to apply a field approximately parallel to the diamond surface. Due to the magnet's shape, there is a trade-off between the applied field strength, and the fields alignment with one of the $\langle111\rangle$ directions. When attempting to increase the field strength beyond 10 mT by bringing the magnet closer, it is not possible to selectively target a specific crystallographic sub-group. An unequal off-axis field is therefore inherently applied to all four crystallographic subgroups at $>10$ mT. 
	
	A spectrum of the ensembles fluorescence under 532 nm excitation is shown in Fig.1(c) highlighting the presence of both the charged N-$V^-$ and the uncharged N-$V^0$ states, and the negligible fluorescence at the chosen probe wavelength of 960 nm. A spectrum of the optically detected fluorescence contrast while sweeping the microwave frequency is shown in Fig.1(d) (an identical spectrum from the same diamond is shown in \cite{El-Ella2019a}), which was measured using a lock-in amplifier and amplitude-modulation of the microwave field with a -5 dBm peak power. A magnetic field at approximately 30 mT is applied at an oblique angle to all $\langle 111 \rangle$ axis which energetically distinguishes all four crystallographic subgroup resonances related to the $\vert0\rangle \leftrightarrow \vert\pm1\rangle$ spin transitions.
	
	P1 electron spin resonances are directly detectable via the N-$V^-$ fluorescence contrast, which occur here between 600 and 850 MHz. These resonances are likely detectable due to a Raman-based cross-relaxation between the N-$V^-$ and P1 electron spins, as discussed in \cite{Purser2019}, and consist of nine peaks based on the hyperfine coupling of the P1's electron ($S{=}1/2$) and quadrupolar nuclear ($I{=}1$) spins: five peaks originating from allowed $\vert m_I, {-} 1/2\rangle{\leftrightarrow}\vert m_I', 1/2\rangle$ transitions, and four low amplitude peaks related to nuclear spin flip-flop and forbidden transitions ($\Delta m_I \neq\{0, 1\}$) \cite{Hall2016}. 
	
	\section{Reflected light polarization}
	The primary experimental signature of the MOVE is witnessed in the measured variation of the reflected light as a function of magnetic field strength, shown in Fig.2(a). Due to the manual rotation mount of the $\lambda/2$ plate, it is difficult to set its absolute optimum angle to obtain a fully \textit{s}-polarised reflection at a given magnetic field. A changing magnetic field will therefore rotate the polarisation around this point, as demonstrated in Fig.2(a) showing a quadratic parabolic dependence of the detected light as the magnetic field strength is linearly increased. This measurement was also performed with a electronic grade diamond with a negligible nitrogen concentration ($<5$ ppb), which did not display any polarisation rotation. 
	
	The dependence on the reflection angle is shown in Fig.2(\textit{c},\textit{d}) at approximately 20 mT. The rotation of the $\lambda/2$ plate maps out a sinusoidal dependence of the PBS output, and highlights its finite contrast ratio by the detected difference never fully reaching zero. A quadratic increase in amplitude at normal incident is measured, while a reversed approximately linear dependence is measured for orthogonal polarised incident light in Fig.3(b) (note that the $\lambda/2$ rotation angle imparts a $2\varphi$ rotation of the incident polarisation). The measured relative change in the detected difference is estimated to represent a rotation in the order of $0.1^\circ$.    

	Given the possibility for optically polarising the N-$V^-$ centres electron spin into its $\vert0\rangle$ ground state \cite{Manson2006}, the overall ensemble magnetisation can be perturbed by continuous optical excitation at a wavelength of 532 nm. This is expected to modify the rotation of the reflected light polarisation by extension of the MOV effect, as demonstrated in Fig.3. This highlights an opposite change in detected reflected light for orthogonally polarised incident light, while the diamond is illuminated using 532 nm. The observed 30 - 40 second time-scale of this rotation is orders of magnitude slower than the $0.1$ $\mu$s-0.1 ms relative lifetimes of the indirect electron transition rates in diamond under an obliquely applied magnetic field to the N-$V^-$ symmetry axis \cite{Tetienne2012}. 
	
	Although the off-axis magnetic field components results in the mixing of the spin eigenstates and thereby degrade the efficiency of optically polarising N-$V^-$ using 532 nm \cite{Tetienne2012}, this mechanism is not expected to be fully quenched. Therefore, the measured slow rotation may reflect the slowed polarisation rate of the N-$V^-$ sub-ensemble, and perhaps by extension the P1 spin ensemble, mediated by their coupling. On the other hand, such slow rates also suggest the possibility of slowly heating under optical excitation, given the opaqueness of the diamond crystal due to the defect density, thereby degrading the overall magnetisation and reducing the reflected light's polarisation rotation.
	
	\begin{figure}
		\includegraphics[scale = 0.7]{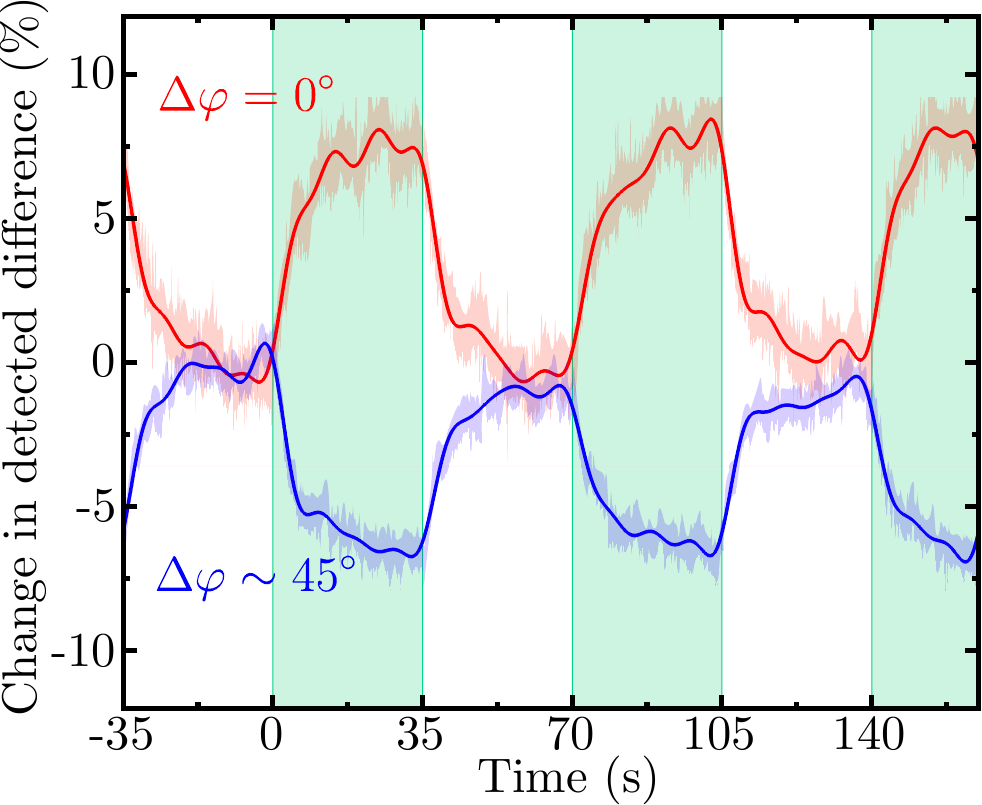}
		\centering
		\caption{Detection of the difference with the cavity's transmission amplitude, gated with a pulsed 532nm at approximately 14 mHz (highlighted by the green regions in the graph). Using two orthogonal incident polarisations of the probe light, an opposite change in the detected amplitude is measured, proving the occurrence of a 532 nm instigated rotation of the reflected light polarisation.}
	\end{figure}

\section{Spectral Characteristics \& Power dependence}
	Insight into the dynamics underlying these observations is obtained by spectrally analysing the reflected light through the FP transmission while pulsing the 532 nm laser. This is shown in Fig.4 for a fully \textit{s}-polarised probe field and a magnetic field at approximately 25 mT. The cavity spectrum is mapped as a function of time while 532 nm is slowly pulsed with a repetition rate of approximately 40 mHz. Each spectrum is fitted with a Voigt function with an approximate Gaussian/Lorentzian (G/L) ratio of 2:1. No significant change in the fitted lineshape is observed, with or without 532 nm illumination, as is evident in the residual plot which displays a slight asymmetry stemming from the imperfect in-coupling of the light into the FP cavity. Only a gradual change occurs for the amplitude and overall linewidth, with no significant change in the G/L ratio.
	\begin{figure}
		\includegraphics[scale = 0.5]{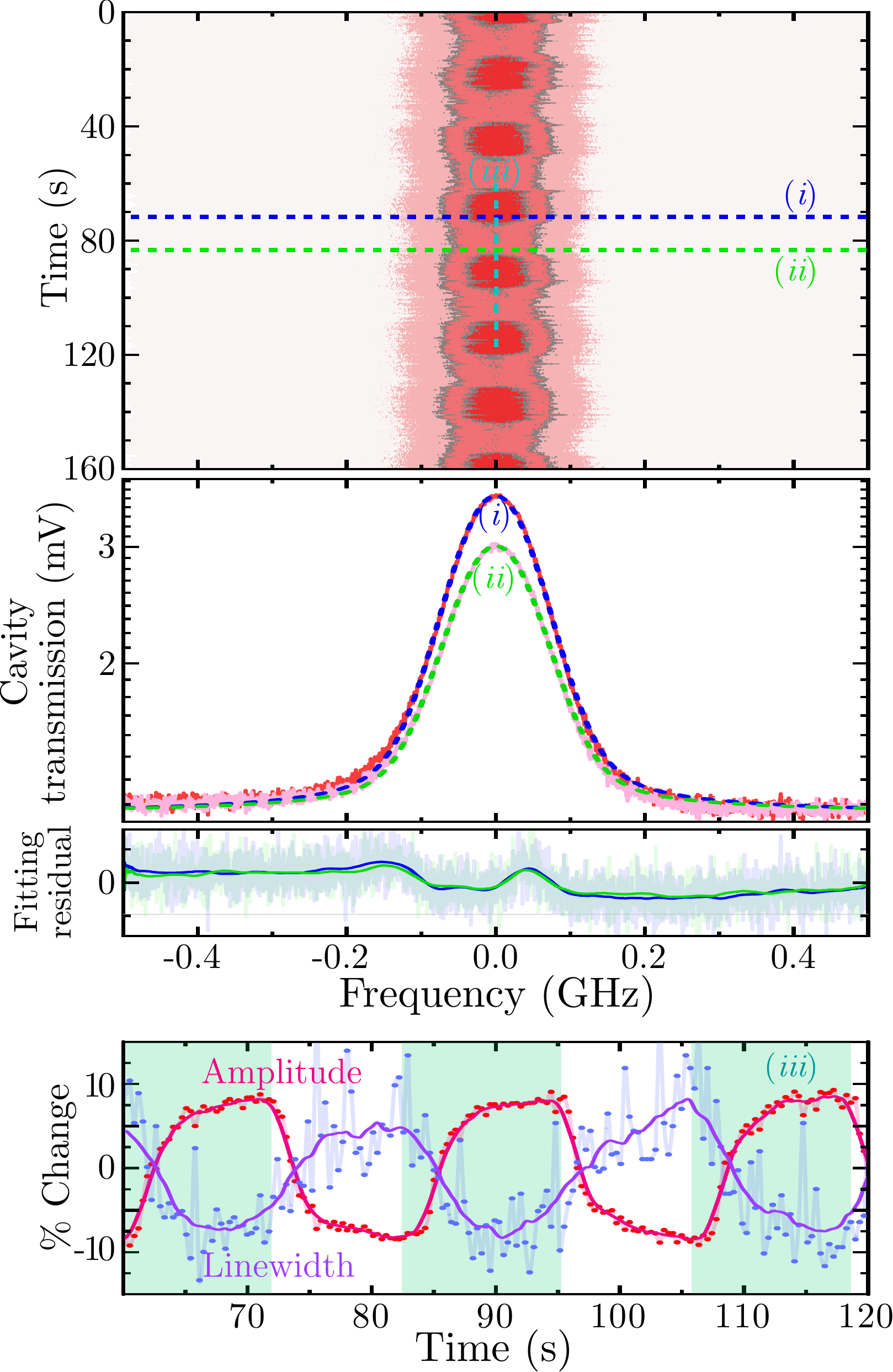}
		\caption{Detected cavity transmission while pulsing 532 nm illumination at $\sim40$ mHz, using a logarithmic colour map. (\textit{i,ii}) Each spectrum is fitted with a convolution of Gaussian and Lorentzian function (known as a Voigt function), and no discernible difference is observed in the lineshape as highlighted by the fitting residual. (\textit{iii}) Distinct opposite oscillations of the transmission peak linewidth and amplitude is detected under 532 nm illumination (highlighted by the green regions in the graph).}
	\end{figure}

	As the sample is illuminated, a narrowing linewidth accompanying a very slight increase of Lorentzian component is measured. No other discernible peaks were detected within the free spectral range of the cavity. The suitability of the Voigt function reflects the observation that the measured lineshape is a convolution of a Lorentzian lineshape inherent to confocal cavity-related transmission, and a Gaussian component associated with stochastic frequency-broadening mechanisms stemming from the crystal, which may also perturb the scattered light polarisation. Such stochastic frequency-broadening mechanisms are usually associated with the excitation of phonon states that impart broad low-energy Brillouin scattering. This is often observed accompanying prominent Brillouin peaks that are Stokes shifted from the central Rayleigh peak, and are sometimes referred to as ``Rayleigh wings" \cite{BARILLE2002969,Bahl2012}. Given the sensitivity of this component and its effect on the linewidth, an increased linewidth is therefore expected for a heated crystal alongside an amplitude increase, as well as a significant change in the G/L ratio. The opposite effect is observed in this case, and no significant change in the G/L ratio is measured, suggesting that no significant heating is occurring, and that the change in detected amplitude is likely related to a slow increase in overall magnetisation at the illuminated spot, which is related to the ensemble spin polarisation.  
	
	Further insight may be obtained when measuring the magnitude and rate of polarisation rotation as a function of 532 nm power. Fig.5 plots both these trends for two different magnetic field strengths, highlighting a non-linear dependence which is well-fit in this case with either quadratic or mono-exponential curves. The dependence of the amplitude change on the magnetic field strength reflects the quadratic dependence of the Voigt effect, however there is no observed magnetic dependence on the rate-of-change trends, which are best fit using bi-exponential curves. 
	
	\begin{figure}
		\includegraphics[scale = 0.8]{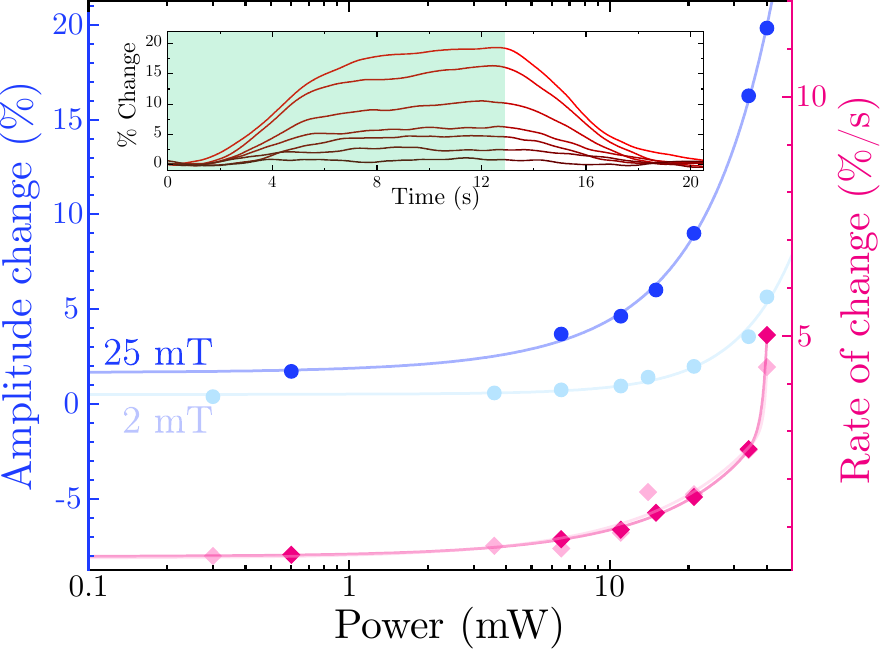}
		\caption{Trend of the maximum amplitude change (circles) and the rate of change (diamonds) as a function of laser power for a $\sim25$ mT and $\sim2$ mT applied magnetic fields (darker and lighter colour, respectively). The inset plot shows the measured data set at 25 mT, with the green region highlighting the 13 s pulse duration.}
	\end{figure}
	
	These trends possibly rule out a heating mechanism, as a near-linear dependence would be expected between demagnetisation and the lasers power, considering Curie's law, and assuming a linear relationship between the laser power and the generated heat. Taking this consideration further, heating-induced demagnetisation would be mitigated by increasing the magnetic field, and an opposite dependence (a lower amplitude change at larger applied magnetic fields) would be expected between the two magnetic-field trends plotted in Fig.5. More significant is the observation of a non-constant rate of change for varying laser powers, which seems independent of the magnetic field strength. This heavily suggests that the basis of this mechanism is not related to dis-allowed and non-spin conserving transitions, whose rates are non-linearly dependent on the strength of off-axis magnetic fields \cite{Tetienne2012}. 
	
	Another conceivable basis for the observations in Fig.3-5 is the modification of the ensembles collective charge-state. Although the charge-state dynamics are still not fully understood, the existence of a   [N-$V^-$]$\rightleftharpoons$[N-$V^0$] equilibrium is experimentally confirmed \cite{Aslam2013,Ji2016}. Furthermore, while the comparatively little understood N-$V^0$ state is predicted to possess a spin 1/2 \cite{Felton2008}, it has yet to be measured. It is therefore reasonable to assume that N-$V^0$ possesses, at the least, a negligible magnetic susceptibility compared to that of N-$V^-$ spin. By extension, the MOV effect would therefore not likely be observable for the N-$V^0$ state, and given the relative abundance of charge-donating P1 defects (N-$V^0$+N$^0$$\rightleftharpoons$N-$V^-$+N$^+$ \cite{Giri2018}), this equilibrium is likely pushed towards the negatively charged state under 532 nm illumination. On this basis, it is postulated that the rotation induced by the 532 nm laser is imparted by increasing the [N-$V^-$] fraction of the ensemble, facilitated by the abundance of P1 charge donors, further increasing the bulk magnetic susceptibility of the diamond membrane, and thereby amplifying the Voigt effect.    
\section{Conclusion}
	The experimental observation presented in this article is, to our knowledge, the first distinct demonstration of modifiable magnetic birefringence in a paramagnetic diamond crystal using 532 nm illumination. Based on the spectral study of the slow dynamics and its power dependencies, it is postulated that the mechanism is based on charge state conversion and its relation to the spin ensembles collective magnetic susceptibility. Further investigation is necessary to accurately assess and quantify these postulations and related magneto-optic constants. This will involve precisely aligning magnetic fields and optical polarisation with respect to the diamonds crystallography, while using probe wavelengths $>$1050 nm to rule out any possible perturbation of the ensemble spins via the N-$V^-$ singlet state. 	
	
	\begin{acknowledgments}
		This work was funded by the Villum Foundation (grant No.17524). U.L.A would like to acknowledge funding from the Danish Research Council through the Sapere Aude project DIMS, and the Danish National Research Foundation (bigQ, DNRF142). We are grateful to Jonas S. Neergaard-Nielsen and Olivier Gobron for helpful discussions.		   
	\end{acknowledgments}
\bibliography{Subm}
\end{document}